\documentclass{IEEEcsmag}
\usepackage{graphicx} % Required for inserting images
\usepackage{listings}
\usepackage{xspace}
\usepackage{hyperref}
\usepackage{amsmath,amsfonts}
\usepackage{listings}
\usepackage{xcolor}
\usepackage{balance}

\definecolor{backcolour}{rgb}{0.95,0.95,0.92}

\newcommand{\nbProjects}{28,193\xspace}
\newcommand{\nbFiles}{640,178\xspace}
\newcommand{\nbArtists}{5,532\xspace}
\newcommand{\nbEditions}{2,705,350\xspace}
\newcommand{\nbSeeds}{2,704,817\xspace}

\lstdefinelanguage{JavaScript}{
  keywords={typeof, new, true, false, catch, function, return, null, catch, switch, var, if, in, while, do, else, case, break, let},
  keywordstyle=\color{magenta}\bfseries,
  backgroundcolor=\color{backcolour},   
  ndkeywords={class, export, boolean, throw, implements, import, this,for},
  ndkeywordstyle=\color{magenta}\bfseries,
  identifierstyle=\color{black},
  sensitive=false,
  comment=[l]{//},
  morecomment=[s]{/*}{*/},
  commentstyle=\color{purple}\ttfamily,
  stringstyle=\color{red}\ttfamily,
  morestring=[b]',
  morestring=[b]"
}
\begin{document}

\title{fx(hash): Salvaging The Source Code of 28k Generative Artworks}

\author{Benoit Baudry, Valeriia Bezklynska, Martin Monperrus}

\author{\textbf{\emph{Université de Montréal, KTH Royal Institute of Technology}}}

\begin{abstract}
    fx(hash), the most radically inclusive generative art NFT platform, went bankrupt and shut down on August 17, 2026. The ownership of its \nbEditions artworks is immutable on the Tezos and Ethereum blockchains, but the source code that generates them is off-chain and in peril. We salvaged all \nbProjects generative programs contributed by \nbArtists international artists and archived them in Software Heritage. Here we explain why NFT generative art is fragile and the steps we took to salvage these software cultural landmarks.
\end{abstract}

\maketitle

\section{Introduction}

Generative art is art that is produced by an autonomous, generative system \cite{galanter2016generative}. Artists who practice generative art engage in a collaboration where their human artistic gesture and the development and operation of this system interact to produce the artwork.

In modern times, generative art is often based on a computer program \cite{baudry2026art}. In this program, the artists encode their personal creative intentions as production rules, and also leave room for emergence through random decisions. In its early forms, this interplay was used by the artist to systematically explore visual variations and curate the output space of the generative program \cite{molnar1975toward}. We refer to an output produced by a generative art program as an edition. 
In a more engaging form of generative art, the artist writes a program and delivers it to the public, e.g., through a web page. In this case, the public explores the output space of the generative program first-hand and chooses which edition they prefer \cite{franceschet2025modern}. Here, the artist does not decide which edition is sold or exhibited, the collectors do. The artist's essential contribution is the program that generates the visual output.

fx(hash) is a major platform for generative art. Launched on the Tezos blockchain in November 2021, it attracted thousands of artists over the years, who have published \nbProjects generative programs. Collectors have acquired \nbEditions editions of these programs, for a total trading volume of about USD 54 million. On fx(hash), an artist uploads the source code of a generative program and sets the number of editions and their price. As long as the maximum number of editions has not been created, collectors continue to explore the space of possible outcomes of the program and acquire the edition they like the most. This edition is recorded on the blockchain as a non-fungible token (NFT), which stores the random seed to recreate the exact edition together with the identity of the owner. The blockchain guarantees ownership: the token cannot be forged or altered, and every sale is publicly traceable. The NFT can be sold to other collectors who wish to acquire an existing edition of a generative artwork.

%% catastrophe, platform closes
On August 17, 2026, fx(hash) went bankrupt and shut down\footnote{Announcement from the founder of fx(hash): \url{https://x.com/fx_hash_/status/2089351038390124661?s=20}}. This closure is a tragedy for digital art. In the generative art ecosystem, fx(hash) was one of the largest public collections of artworks whose medium is code, as opposed to other platforms that accept only images or videos. For the history of NFT art, fx(hash) embodied radical inclusivity \cite{estorick} because its content was not filtered by any art curator. This true openness nurtured a vibrant community of artists and collectors and supported artistic experimentation \cite{fxhashcollectors}. fx(hash) is not an isolated case: many other NFT platforms had to cease their activities for economic reasons, including \href{https://www.foundation.app/?new=true}{Foundation}, which closed in April 2026.

%% our goal
Our goal is to salvage this unique artistic landmark. The major challenge is that the blockchain only stores seeds and ownership. The source code is stored off-chain and can be permanently lost: in other words, the artwork may not survive a platform bankruptcy. When off-chain storage disappears, the tokens on-chain remain but point to nothing, and no edition can be re-executed and displayed. 

To prevent this catastrophic loss, we planned and performed the first complete, long-term preservation of fx(hash). We salvaged the source code of all \nbProjects programs, together with their dependencies and their \nbSeeds seeds. We archived the code in Software Heritage, an archival platform supported by UNESCO, whose lifetime does not depend on any art market or private investor. 

\begin{figure}[htp]
    \centering
    \includegraphics[width=\columnwidth]{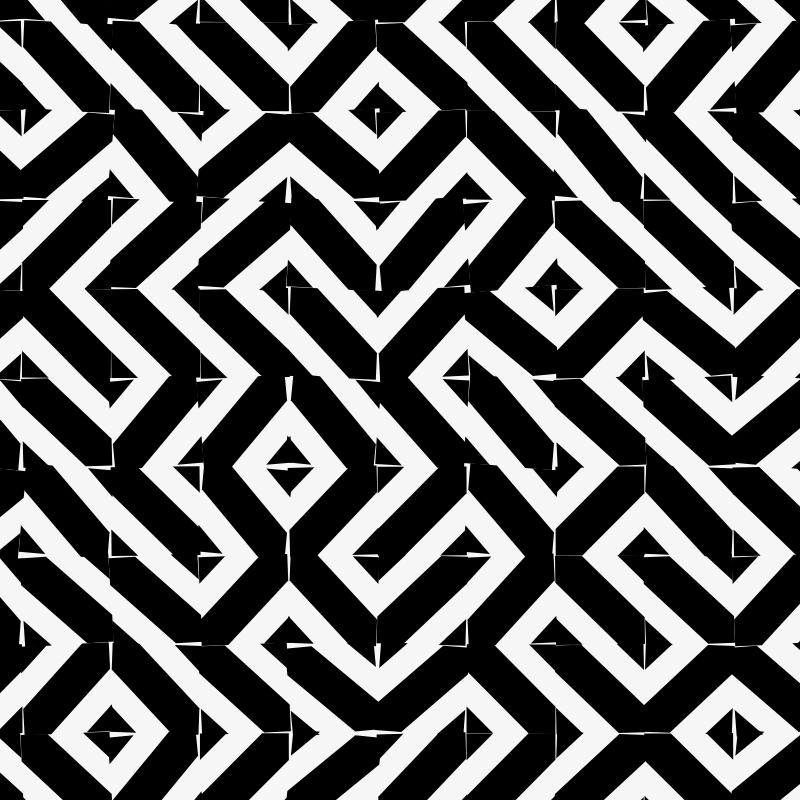}
    \caption{\emph{CENTURY-XXX-METAMOLNAR(S) \#3}, by \href{https://reas.com/}{Casey Reas}. We have archived its source code on Software Heritage \href{https://archive.softwareheritage.org/browse/origin/directory/?origin_url=https://fxhash-preservation.rethread.art/fxhash-1.0-tezos/2021/CENTURY-XXX-METAMOLNAR-S-__4558.git}{swh:1:dir:67e78215c2dac7ef3130eb7d3c606d6c9511a634}.
%https://killedbyapixel.github.io/fxhashArchive/#/gentk/KT1KEa8z6vWXDJrVqtMrAeDVzsvxat3kHaCE/239644?p=4558&s=century-xxx-metamolnar(s)&i=3
    }
    \label{fig:reas}
\end{figure}

\begin{figure}[htp]
    \centering
    \includegraphics[width=\columnwidth]{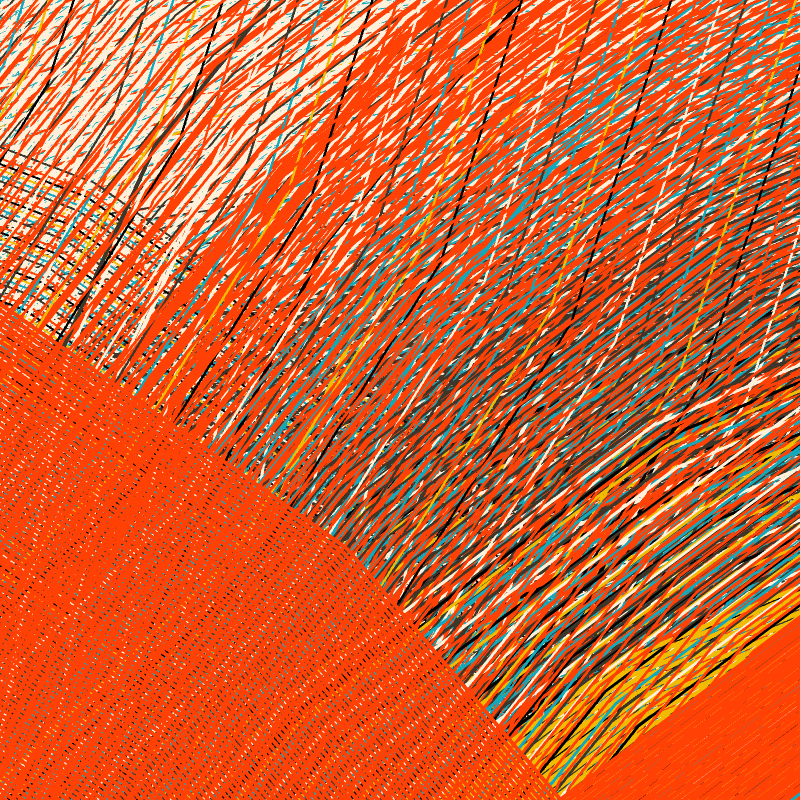}
    \caption{\emph{A Stitch is Fine}, by \href{https://linktr.ee/lisaorthstudio}{Lisa Orth}. We have archived its source code on Software Heritage \href{https://archive.softwareheritage.org/browse/origin/directory/?origin_url=https://fxhash-preservation.rethread.art/fxhash-1.0-tezos/2021/A-Stitch-Is-Fine__3112.git}{swh:1:dir:5e6eb4e84492291da22ab83875f6b7d871b7f2ea}.
%https://killedbyapixel.github.io/fxhashArchive/#/gentk/KT1KEa8z6vWXDJrVqtMrAeDVzsvxat3kHaCE/152167?p=3112&s=a-stitch-is-fine&i=13
    }
    \label{fig:stitch}
\end{figure}

The salvage has been a success. The \nbProjects salvaged programs amount to 121\,GB of material, in \nbFiles files. 
These salvaged programs cover a wide variety of styles and visual concepts. For example, \autoref{fig:reas} shows a black-and-white homage to the generative art pioneer Vera Molnar. This is one of the 13 artworks contributed to fx(hash) by artist and scholar Casey Reas. The generative program is about 100 lines of JavaScript that specifies how to assemble and randomly tilt the tiles on a grid, whose resolution is randomly chosen at runtime.
\autoref{fig:stitch} shows a different style of artwork by artist Lisa Orth, whose source code is shown in \autoref{lst:stitch}. This program defines a palette of colors and then randomly selects in this palette to draw a composition of arcs on the canvas. The generative programs for these two artworks, like all the others, are now preserved in Software Heritage.

% \begin{figure}[ht]
%      \centering
%      \footnotesize
\begin{lstlisting}[language=JavaScript, escapechar=!, linewidth=\columnwidth,caption=Generative source code for the artwork in \autoref{fig:stitch},label={lst:stitch}]
//"A Stitch Is Fine" by Lisa Orth  @lisaorthstudio
let colors = ["#3C362B", "#FEF0D9", "#FF4106", 
  "#000000", "#F0BB01", "#12A9BD"];
function setup() {
 m = min(windowWidth, windowHeight);
 createCanvas(m, m, WEBGL);
 let st = fxrand() * 4 + 2
 translate(m / st, m / st);
 scale(fxrand() * 1.5 + 1.2); rotate(10);
 background("#FFFFFF");
 stitch();
}
function stitch() {
 function colourSelect(rand) {
  if (rand < 1) {
   return color("#12A9BD");
  } else if (1 < rand && rand < 2) {
   return color("#3C362B");
  } else if (2 < rand && rand < 3) {
   return color("#FEF0D9");
  } else if (3 < rand && rand < 4) {
   return color("#FF4106");
  } else if (4 < rand && rand < 5) {
   return color("#F0BB01");
}}
 let part = int(fxrand() * 29 + 10);
 let ms = m / part;
 translate(fxrand()* 49 - 50, fxrand() * 99 - 100);
 scale(fxrand() * 1.5 + 1.4);
 translate(fxrand()* 49 - 50, fxrand() * 49 - 50);
 background(colourSelect(fxrand() * 5));
 for (let i = 0; i < fxrand() * part; i++) {
  for (let j = fxrand() * 10; j < part * 10; j++) {
   let x = i * ms + ms / 2;
   let y = j * ms + ms / 2;
   let c = fxrand() * 19 + 5 * ms * PI;
   let osx = fxrand() *9-10*c*fxrand()*1.5+0.25;
   let osy = osx * fxrand() * 2 + 0.5;
   strokeWeight(fxrand() * 4 + 0.75);
   stroke(colourSelect(fxrand() * 5));
   duplCirc(x, y, c);
   stroke(colourSelect(fxrand() * 5));
   duplCirc(x + osx, y + osy, c);
}}
 function duplCirc(x, y, c) {
  noFill();
  let crv = int(fxrand() * 4.2);
  let cnt = int(c * fxrand() * 0.42 + 0.12);
  for (let i = 0; i < cnt; i++) {
   let cc = map(i, 0, cnt, c, 0);
   if (crv == 0) {
    stroke(colors[i % colors.length]); }
    circle(x, y, cc);
} } }
\end{lstlisting}
% \caption{Source code for \emph{A Stitch is Fine}.}
% \label{fig:generative-art-code}
% \end{figure}

\section{Preserving Code-Based Art}

Software opens fantastic opportunities to amplify creativity, letting artists, designers, and composers experiment with novel media, novel production techniques and novel interactions. 
Yet, software artworks are also highly vulnerable to rot and decay. Rapid software evolution \cite{engel2014reading}, hardware obsolescence, insufficient documentation \cite{stricot2018open}, and ephemeral third-party dependencies \cite{wang2023nfts} all threaten their longevity and reproducibility. 

As these software-based artworks have become a significant part of our cultural heritage, museums, artist-run collectives and art studios have started to work on dedicated techniques for their long-term preservation. 
For example, hardware emulation has been used in several contexts to preserve artworks that depend on hardware that is no longer available. Rhizome is an international organization that provides emulation as a service for museums and galleries \cite{espenschied2016exhibiting}.

Many generative artworks also depend on the web to dynamically load resources, such as images or data. In this case, the Internet Archive's \href{https://web.archive.org/}{Wayback Machine} or \href{https://rhizome.org/editorial/2020/jun/11/introducing-conifer/}{Rhizome's Conifer} provide historical copies of publicly accessible Web resources so that artworks keep working. Sch\"utze recently documented how he used network control to restore two internet-based artworks from the ZKM collection \cite{Schutze26}. 

%code-based testing and migration
Several preservation efforts have investigated code migration for generative artworks. A particularly important moment for these operations was the shutdown of Adobe's Flash, which was a favorite technology for early generative and net artists. After the Flash shutdown, many artworks were migrated to native browser technology such as HTML and JavaScript \cite{mladentseva2022responding}. 

This previous work investigates strategies to preserve the ability to execute generative art, assuming the availability of the code (source or binary) and all the assets required to run the artwork. NFT artworks introduce specific challenges for preservation: the artwork is often composed of many components (metadata, media assets, code) that are scattered off-chain \cite{brizi,Balduf2022DudeWMA} and calls for dedicated preservation.

\section{NFT Architecture}

An NFT artwork is composed of several components.
A typical workflow to create a generative art NFT goes as follows. First, an artist publishes an artwork on the NFT platform. In the context of fx(hash), the artist uploads the source code, libraries and all assets needed to run a generative art program. For example, Lisa Orth uploaded five files that can generate various editions of \emph{A Stitch is Fine} (see \autoref{fig:stitch}): an HTML entry point, a CSS style sheet, the file \texttt{sketch.js} that contains the actual generative code (\autoref{lst:stitch}), as well as two third-party libraries, \texttt{p5.js} and \texttt{p5.sound.js}. 
This source material is the \emph{art code} and is stored outside the blockchain. 

Second, art collectors visit the NFT platform and browse the output space of the generative program. As the code includes random decisions, every execution produces a unique edition of the artwork. Once they find one that they like, they create a token on the blockchain that records the random seed to regenerate this edition. This operation is referred to as ``minting'' an NFT on the blockchain, and the seed that is saved is called the \emph{art token}, which is non-fungible. 
Third, when the collector wants to see their artworks, they go to the platform, which retrieves the token from the blockchain, the code from elsewhere, and recreates the exact edition variant that the collector selected, with the same visual and audio properties.

\begin{figure}[htp]
    \centering
    \includegraphics[width=\columnwidth]{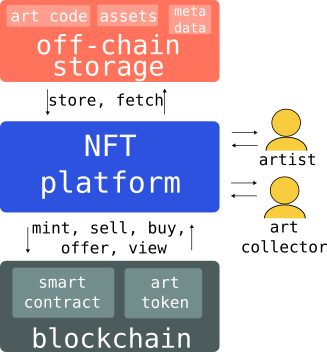}
    \caption{NFT platform to publish, curate and trade generative artworks.}
    \label{fig:nft-platform}
\end{figure}

Figure~\ref{fig:nft-platform} depicts those three layers involved in the generation, distribution and storage of an NFT artwork: the blockchain, the off-chain storage, and the NFT platform that connects them to artists and collectors.

\textbf{Blockchain layer}
The blockchain stores the smart contracts and the tokens \cite{wang2021nft}. A smart contract defines the rules of a collection: the maximum number of editions, the price, and how the revenue of primary and secondary sales is split between the artist, the platform, and other parties. A token records the ownership of a specific variant. Storage on a blockchain is expensive, so a token usually holds a small amount of data, and not the entire artwork: the owner's address, and a pointer to content stored off-chain.
fx(hash) first deployed its contracts on the Tezos blockchain and in 2023 expanded to Ethereum. The fx(hash) art tokens store the owner's address, the random seed to recreate the specific minted edition and a URI that points to the art code that is stored off-chain.

\emph{Preservation.} The blockchain layer is the most durable of the three layers. Its data is replicated on every node of the distributed network and is immutable. This means that the content of a token cannot be modified after the mint. As long as the Tezos and Ethereum networks run, the ownership, the transaction history, and the seeds of the \nbEditions fx(hash) editions remain available. Preserving this layer consists of reading the seeds and pointers from the chain. 

\textbf{Off-chain storage layer}
The generative program, its dependencies (e.g., third-party libraries such as p5.js, fonts, audio snippets), and the preview thumbnail images are too large to be stored on a blockchain. NFT platforms store them off-chain and include an address to this off-chain material in the art token. fx(hash) stores this pointer in the \texttt{generative\_uri} field of each on-chain token. Most NFTs use IPFS (InterPlanetary File System), a protocol for a peer-to-peer file system in which a file is addressed by the hash of its content \cite{benet2014ipfs}. A content address guarantees integrity: the bytes retrieved for a given address are the bytes the artist uploaded. A content address does not guarantee availability: a file stays on IPFS only as long as at least one node stores it, an operation called pinning.

\emph{Preservation.} The off-chain storage layer is the most fragile one. It relies on persisting all the IPFS content and most NFT platforms use a third-party service to manage this IPFS layer. When fx(hash) stops paying the third-party pinning service, the content of the IPFS addresses disappears while the on-chain pointers remain. Wang et al. \cite{wang2023nfts} measured this failure mode at scale and found that many NFTs already point to content that can no longer be retrieved. Preserving this layer requires copying the files before they disappear and storing them in an archive whose lifetime does not depend on a commercial platform.

\textbf{Platform layer}
The NFT platform is the software that artists and collectors interact with. It provides a web front-end to upload a project, configure its contract, mint, buy, sell, and display editions. 
Essentially, the platform defines the execution protocol between the program and its host page. The fx(hash) platform is a web application that can fetch a seed and a URI pointing to the art code from the blockchain. Then, the \texttt{fxpreview()} function runs the art code with the random seed to display a preview.

\emph{Preservation.} The platform layer is centralized and its lifetime is the same as the lifetime of the company that operates the platform. After the shutdown of fx(hash), the \href{https://www.fxhash.xyz/}{fxhash.xyz} front-end, the API, and the preview service are no longer available. Two elements of this layer must be documented to keep the artworks executable.
First, one must store the mapping between projects, tokens, seeds, and storage addresses.
Second, one must reverse engineer the execution protocol, i.e., the URL parameters and the functions that a program expects from its host page in order to be executed.
Without this information, the source code might be preserved but no edition can be reproduced from it with proper execution.

\section{The fx(hash) Preservation Project}

\begin{table*}[thp]
    \centering
    \caption{Reference smart contracts of fx(hash). Tezos figures come from TzKT. Addresses are truncated to their first 10 characters. Generative artworks are the entries currently in the registry of an issuer contract, each counted once, in the latest registry where it appears. All projects of issuer v1 were copied to issuer v2. Each token is one minted edition, with one specific seed. On Ethereum and Base, projects are the contracts listed in the registry of each factory, and tokens were read from these project contracts; transactions are those sent to the factories, the deployment date is that of the first factory, and EVM figures come from the chain.}
    \label{tab:contracts}
    \footnotesize
    \begin{tabular}{llllrrr}
        \hline
        \textbf{Contract} & \textbf{Role} & \textbf{Address} & \textbf{Deployed} & \textbf{Transactions} & \textbf{Artworks} & \textbf{Tokens (seeds)} \\
        \hline
        \multicolumn{7}{l}{\textit{Tezos}} \\
        Issuer v1 & project registry & \texttt{KT1XCoGnfu} & 2022-01-04 & 1,167,397 & (ported to v2) & \\
        GENTK v1 & minted tokens & \texttt{KT1KEa8z6v} & 2021-11-03 & 3,006,555 & & 589,146 \\
        Issuer v2 & project registry & \texttt{KT1BJC12dG} & 2022-04-14 & 4,245,921 & 22,649 & \\
        GENTK v2 & minted tokens & \texttt{KT1U6EHmNx} & 2022-04-14 & 4,031,309 & & 1,030,141 \\
        Issuer v3 & project registry & \texttt{KT1Xpmp15K} & 2023-03-18 & 1,115,338 & 4,783 & \\
        GENTK v3 & minted tokens & \texttt{KT1EfsNuqw} & 2023-03-18 & 663,493 & & 226,346 \\
        \hline
        \multicolumn{7}{l}{\textit{Ethereum and Base}} \\
        Ethereum & 2x project registry & & 2023-12-11 & 140 & 143 & 8,510 \\
        Base & 4x project registry & & 2024-02-19 & 1,886 & 797 & 851,207 \\
        \hline
        \textbf{Total} & & & & 14,232,039 & 28,372 & 2,705,350 \\
        \hline
    \end{tabular}
\end{table*}

The fx(hash) smart contracts deployed on Tezos and Ethereum store the reference for what exists at the time the platform closed. From these contracts we can determine exactly which art code projects were published, by whom, when, and which editions were minted with which seed. \autoref{tab:contracts} lists these contracts and the number of artworks they hold.

Between 2021 and 2023, fx(hash) deployed three versions of its smart contract on Tezos (v1, v2, v3). Each contract version is composed of two types of contracts. 
The issuer contracts hold the registry of art code projects (Issuers v1, v2, v3). In issuers v1 and v2, each contract holds one IPFS identifier per artwork that points to a metadata document that includes the address of the code. 
Issuer v3 stores the address of the code on chain, in a separate entry. 
The GENTK contracts hold the minted tokens, i.e., the editions: 589,146, 1,030,141, and 226,346 tokens for versions 1, 2, and 3. 
On Ethereum and Base, every artwork has its own token contract, deployed by a factory contract. fx(hash) used two factories on Ethereum and four on Base. %Of the 940 EVM artworks registered by these 4 factories.

\textbf{Collecting the fx(hash) material for preservation} 

From the smart contracts we can collect the two essential pieces of data to preserve the fx(hash) artworks: the code for the \nbProjects artworks\footnote{\autoref{tab:contracts} counts 28,372 projects. The difference with the \nbProjects we collect comes from 176 EVM projects that were never minted, and 3 Ethereum projects whose code was only stored on the fx(hash) servers.}, as well as the \nbEditions seeds that have been minted by the collectors.

%code
The code uploaded by the artists is packaged in one folder per artwork and usually consists of a collection of files: an \texttt{index.html} entry point, JavaScript source code files, third-party libraries, media assets such as sound files, shaders, fonts, and additional files such as a license.
From the smart contracts, we retrieved the addresses of the artworks. For 27,604 of the \nbProjects artworks, this is an IPFS address. We retrieved the code of these artworks through a public IPFS gateway. The 589 other projects store their code directly on chain, in ONCHFS, the on-chain file system of fx(hash). % ONCHFS 
The full collection of artworks' source code consists of \nbFiles files, totaling 121\,GB. JavaScript source code is 35\,GB, and the rest is composed of assets such as image and sound files, 3D textures, or data files.

% seeds
The second piece of data for preservation is the set of seeds recorded at minting, and the parameters chosen by the collector when the artwork allowed them. The code without the seeds produces new random variants. The seeds without the code produce nothing. For example, the seed to generate the specific edition of \emph{A Stitch is Fine} shown in \autoref{fig:stitch} is \texttt{ooMxFjhAb95WXdYwBHg4MZjsw41R8KTjQb2rqraF5\\ozB7Yv3Qje}. 
With this seed, every random number in the code of \emph{A Stitch is Fine} is regenerated exactly as in the specific execution that the collector decided to mint, with the exact series of color choices and arc compositions.
We collected the seeds from the chains and from the token metadata. On Ethereum and Base, we read the 859,717 seeds directly from the token contracts. On Tezos, the seed of 1,842,794 tokens is in the token's metadata document, which we fetched from IPFS and checked against the hash stored on chain. We published the \nbSeeds seeds on Zenodo: \url{https://zenodo.org/records/24141706}. 533 tokens of the third Tezos contract have no seed: the contract records them as never assigned, and fx(hash) never published their metadata.

\textbf{Long-term archive of fx(hash) art code}

To prepare the code for archiving, we converted each artwork folder into a git repository. A single commit holds the complete folder. The commit's author is the artist and its date is the creation of the project on chain. 
%For the 169 artists who set no name at all, we use their wallet address.
The e-mail field contains the wallet address of the artist, which links each repository to an on-chain identity. 
% For 42 projects, the artists had uploaded the \texttt{.git} directory of their own repository to IPFS, and we kept this original development history. % cool but not necessary in the paper
We published the \nbProjects repositories as a git forge at \url{https://fxhash-preservation.rethread.art/}. 

To verify that the archived code is the code that each token points to, we recompute the content address of every repository, its IPFS address or its ONCHFS root, and compare it with the address stored on chain. Since a content address is the hash of the content, a match proves that the files are identical. 27,782 of the \nbProjects repositories rebuild their on-chain address exactly. 46 more rebuild it once the \texttt{.git} folder that the artist uploaded is put back. 47 hold the same files but the checksum does not match because of file permissions or empty folders. The 318 remaining projects cannot be checked, because their contract stores only an https URL, with no checksum of their content.

For long-term preservation, we rely on Software Heritage \cite{dicosmo2017swh}, a non-profit archive of publicly available source code whose lifetime does not depend on any NFT platform or art market. We pushed the code of all fx(hash) artworks to Software Heritage.
Once a repository is archived, Software Heritage assigns it an intrinsic identifier (SWHID) computed from its content, which remains valid as long as Software Heritage stays up and running.
Thanks to our preservation project, the list of \nbProjects source code repositories contributed by \nbArtists artists on fx(hash) is now safe and can be browsed:
\url{https://archive.softwareheritage.org/browse/search/?q=fxhash-preservation.rethread.art}

Mission accomplished: we salvaged the entire code of all \nbProjects fx(hash) artworks and stored it on a long-term archival platform.

\section{Open Challenges}

With the long-term archiving of fx(hash) generative art source code, we have achieved the most critical step for the preservation and recreation of fx(hash) artworks. In this section, we discuss some of the open challenges to preserve the full experience of the fx(hash) platform.

\textbf{The Obsolescence of Software Dependencies}

Each of the \nbProjects code repositories that we archived is complete. They include all the code, libraries, media files, and configuration files that were uploaded by the artist in the first place on fx(hash). However, the artworks may require a number of dependencies that have become or will become obsolete. Examples of such dependencies include browser features (e.g., WebGL or sound APIs that evolve regularly), hardware (e.g., evolving GPU APIs can break shaders) or third-party services (e.g., a remote API serving real-time events, such as train traffic messages or short messages on a specific topic, which can be terminated). 

%1,115 of the 27,975 artworks we scanned (4\%)
Looking only at the HTML entry point of the artworks, we found 1,115 artworks that load resources from other servers: JavaScript libraries from cdnjs (423 artworks), jsDelivr (227) and unpkg (132), sketches from openprocessing.org (253), and fonts from Google Fonts (about 60). These artworks run only as long as these servers keep serving the same files. The obsolescence can already be observed: cdn.rawgit.com, used by 11 artworks, has shut down. 

To address this challenge, the next steps must focus on the systematic documentation and archiving of these implicit, fragile dependencies. This requires executing each generative program. If the program crashes, log analysis may reveal which dependencies are missing and should be retrieved. If the execution is successful, dedicated monitoring solutions provide the exact set of features and APIs that are necessary. Based on this documentation phase, it is possible to collect the dependencies and archive them with Software Heritage or the Internet Archive for future reconstruction or emulation. 

\textbf{Preserving the fx(hash) Distribution Community}

The second important aspect of fx(hash) is its active, diverse community. Members of this community enjoy browsing the artworks on the platform, looking at specific collections of artworks, collecting, selling, and trading.
The extinction of the platform was a shock for the community. Artists lost access to their work, collectors lost the ability to browse their collection and trade, and generative art enthusiasts lost the ability to navigate this large gallery of international generative art.

With the code, the seeds and the blockchain, it is possible to recreate an online gallery space where the community can meet and enjoy the art.
\href{https://killedbyapixel.github.io/fxhashArchive/}{fxhashArchive} or \href{https://whitehash.m3000.io}{whitehash} are early examples of such community platforms. In the long term, the challenges of creating an alternative platform are socio-technical: recreating and maintaining the platform on top of the code archived in Software Heritage is technical, while establishing a healthy community and supporting trade of these generative artworks poses major social challenges.

\section{Conclusion}

fx(hash) (2021-2026) has been a major arena for radical exploration in generative art. The platform's inclusivity attracted a distinctive community of generative artists, art enthusiasts and art collectors. As such, the \nbProjects generative programs published on fx(hash) are now part of our cultural and digital heritage. We have designed and executed the preservation of this artistic heritage.

% target
Thanks to the long-term archival of the fx(hash) artworks, art conservators, art students, media archaeologists and software engineers will be able to study this special moment of digital art history. They will be able to analyze the artworks' source code, execute it, and, why not, remix it to create even more generative artworks.

\section{Acknowledgments}
We thank \href{https://ciphrd.com/}{ciphrd} for creating fx(hash). We thank Software Heritage for providing a long-term archive for code. 
This research was undertaken thanks to funding from the Canada Research Chairs Program, NSERC and from Mitacs Globalink.

\bibliographystyle{ieeetr}
\bibliography{biblio}

\textbf{Benoit Baudry} is a Professor of Software Engineering at the Université de Montréal, Canada. He received his PhD in 2003 from the University of Rennes and was a research scientist at INRIA (France) from 2004 to 2017 and a Professor at KTH (Sweden) from 2017 to 2023. His research in software engineering focuses on software supply chains and automatic diversification. He also turns code into visual art and counsels for digital art restoration. Read about Benoit at \url{https://softwarediversity.eu}.

\textbf{Valeriia Bezklynska} is a Bachelor's student at the National Technical University of Ukraine, specializing in software engineering. Valeriia was an intern at Université de Montréal when she contributed to the fx(hash) generative art salvage operation.

\textbf{Martin Monperrus} is a Professor of Software Technology at KTH Royal Institute of Technology in Stockholm, Sweden.
Martin is an IEEE Fellow, ``for pioneering machine learning use in code assistance and program repair, impacting millions of developers.''
Read about Martin at \url{https://monperrus.net/martin}.

\end{document}